\documentclass[usenatbib,useAMS,preprint,psfig2]{mn2e}
\usepackage{times}
\usepackage{amsmath}
\usepackage{amssymb,amsfonts}
\usepackage{graphicx}
\usepackage{epstopdf}

\begin{document}
\title[On the origin of the drifting subpulses in radio pulsars]
{On the origin of the drifting subpulses in radio pulsars}
\author[G. Gogoberidze et. al.]{G. Gogoberidze$^{1}$\thanks{E-mail:
gogober@geo.net.ge},  G. Z. Machabeli$^{1}$, D. B. Melrose$^{2}$ and Q. Luo$^{2}$
\\
$^{1}$Center for Plasma Astrophysics, Abastumani Astrophysical
Observatory, Al. Kazbegi Avenue 2a, 0160
Tbilisi, Georgia\\
$^{2}$School of Physics, University of Sydney, NSW 2006,
Australia}

\date{}

%\pagerange{\pageref{firstpage}--\pageref{lastpage}} 
\pubyear{2004}

\maketitle

\label{firstpage}

\begin{abstract}
A model for the main observational characteristics of the radio emission of pulsars with well organized drifting subpulses is presented. We propose that drifting subpulses result from the modulation of the radio emission mechanism due to long-wavelength drift waves in the magnetosphere. The drift waves are generated at shorter wavelengths, and their nonlinear evolution favours accumulation in a specific azimuthal eigenmode with an integral number, $m$, of nodes encircling the magnetic pole. The electric field of the drift waves is along the magnetic field lines, and this modulates the distribution for particles and hence the radio emission mechanism. The ratio of the frequency of the eigenmode to the rotation frequency of the star is insensitive to the magnetic field strength and the period of rotation, and is of order unity. The period, $P_3$, of the drifting subpulses is attributed to the mismatch between this frequency and the nearest harmonic of the rotation frequency of the star.
\end{abstract}


\section{Introduction}

The observational characteristics of drifting subpulses in pulsars were studied in detail in the 1970s \citep{M71,THHM71,RCB74,MTH75,R83,R86} and have attracted considerable interest in recent years \citep{Retal02,vLetal02,vLetal03,Getal04}. Earlier ideas on the polarization-angle rotation have not been confirmed, and switches between orthogonal polarizations are found to play a central role in apparent depolarization \citep{Retal02}. A detailed study of PSR 0809+74 \citep{vLetal02,vLetal03} has clarified the interplay between subpulse drifting and nulling. These and other investigations have led to renewed interest in the interpretation of subpulse drifting. The basic interpretation of subpulse drift involves a pattern that drifts around the magnetic equator such that an individual subpulse reappears in the pulse window after a characteristic rotation time, $P_3$. A generic model, referred to as the carousel model, involves the drifting pattern being identified with plasma columns spaced periodically around a circle that rotates relative to the stellar surface about the magnetic pole. The number of columns and the period of the pattern rotation have been determined for several pulsars \citep{DR99,DR01,vLetal02,vLetal03}, and this number is of order 20. The patterns implicit in the carousel model may be interpreted in a variety of ways. One interpretation is that there are actual plasma columns, for example formed through sparking on the stellar surface, as envisaged in the model of \cite{RS75} and reconsidered recently by \cite{GMG03}. Another interpretation is in terms of a magnetic field with a high order multipole component \citep{AK02}. More generally, the distribution of plasma, magnetic field and other relevant quantities at a particular radial distance from the star can be expressed as a sum of spherical harmonic components, $\ell,m$, with $\ell=0,1,2.\ldots$, and $-\ell\le m\le \ell$, and an interpretation of the carousel model is that a specific such harmonic with $m\sim20$ dominates all other and leads to the inferred pattern of $m$ nodes or anti-nodes around a circle at fixed magnetic latitude. In one specific such model the pattern is interpreted in terms of resonant cavity modes \citep{Y04} in a model in which the polar-cap region acts as the cavity. Another possible class of models involves interpreting the spherical harmonic components as long-wavelength plasma wave modes such that the ratio of the wavelength is not negligible in comparison with the radius of the star. This is familiar in helioseismology, in which the waves are sound-like modes within the star; the spectrum of variations in the observed radiation from a star has sharp peaks that are associated with the natural modes of the star, described by specific $l,m$ values. Such a model was proposed by \cite{CR04}, who invoked a spherical mode with large $\ell\sim70$ and $m=0$. A point made by \cite{CR04} is that the frequency of the wave can play an important role in the interpretation of $P_3$. In the conventional carousel model the columns are assumed stationary in a corotating frame, and there is no counterpart of the frequency of oscillation of the wave. However, a model based on internal oscillations of the star does not lead to a natural way of modulating the radio emission, which is thought to originate well above the surface of the star, and it is unclear why the inferred pattern should be fixed relative to the magnetic axis rather than to the rotational axis.

In the present paper we propose a model for drifting subpulses that is based on long-wavelength drift waves that form a period pattern around a circle centred on the magnetic pole. As shown by \citet{KMM91a,KMM91b}, drift waves propagating across the field lines can be generated in the electron-positron plasma of the pulsar magnetosphere through a resonant instability driven by the oppositely directed drift motions of electrons and positrons. Drift wave have their wave vector, ${\bf k}$ predominantly in the plane defined by the curvature drift, which is almost exactly perpendicular to the magnetic field lines. When generated, the waves have a relatively short wavelength compared with the dimensions of the magnetosphere. Drift waves have a low group velocity, so that the wave energy tends to build up locally rather than propagating away, and their growth is limited by nonlinear processes which redistribute the wave energy to other scales. In particular, induced scattering leads to an energy transfer to lower frequencies which correspond to longer wavelengths. In close analogy with the formation of a Langmuir condensate \citep{G83} in an unmagnetized nonrelativistic plasma, the wave energy accumulates in the long-wavelength region of phase space where there is no effective dissipation mechanism and where the propagation speed of the wave energy is very slow. For wavelengths of order the transverse scale of magnetosphere, the treatment of the drift waves as plane waves becomes inappropriate, and the geometry of the magnetosphere needs to be taken into account. A treatment in terms of spherical harmonic eigenmodes with specific $\ell,m$ is then appropriate: an assumption of strict spherical geometry would be artificial, but the description of the azimuthal distribution in terms of eigenmodes varying as $\exp(im\phi)$ should remain valid in a more realistic model, and it is only the $m$ values that are directly relevant in the model proposed here. Long-wavelength drift waves are then represented in terms of eigenmodes with specific azimuthal wave numbers corresponding to an integer number, $m$, of nodes around a circle centred on the magnetic pole. We argue that the nonlinear processes that transfer the waves to long wavelengths should result in a specific $m$ value dominating. The frequency of these drift waves can be of the order of pulsar rotation frequency, $\Omega=2\pi/P$, where $P$ is the pulsar period. Then, as in the model of \cite{CR04}, the time $P_3$ is attributed to a mismatch between the wave frequency and a harmonic of the rotation frequency. 

There are several attractive feature of the drift-wave model for drifting subpulses. One attractive feature is that drift waves provide a natural way of modulating the radio emission. Specifically, the electric field in the waves is primarily along the magnetic field, and this modulates the acceleration of particles and the form of their distribution function. Although there is no consensus on the pulsar radio emission mechanism, the most widely favoured mechanisms involve waves growing due to a Cerenkov instability driven by free energy associated with an increasing portion of the particle momentum distribution function. This function, and hence the instability, is modulated by the electric field in the drift waves. At a specific height and magnetic latitude, the emission should then have $m$ peaks in azimuth, and should vary temporally at the wave frequency, so that $P_3$ results from the mismatch of this frequency and the nearest harmonic of the rotation frequency. Another attractive feature is that drift waves are intrinsic to the magnetosphere. The pattern of drift waves, described by the distribution among the spherical modes $\ell,m$, is likely to change slowly with radial distance from the star, with the drift pattern at a specific observational frequency determined by the dominant $\ell,m$ mode at the radius at which that frequency is emitted (assuming some form of radius-to-frequency mapping). This avoids the constraint, implicit in the models based on sparking \citep{RS75} or stellar oscillations \citep{CR04}, that the pattern of drifting subpulses be tied to surface phenomena on the star, for which there is no direct evidence.

The paper is organized as follows. The generation of the drift waves and their nonlinear evolution are discussed in section 2. Our proposed model for drifting subpulses is presented in section 3. The conclusions are summarized in section 4.

\section{Generation of drift waves}

Drift waves grow due to an instability driven by the curvature drift, which is proportional to the Lorentz factor of the particles. Although the secondary pairs are much more numerous than the beam particles, they have much smaller Lorentz factors. Both the dispersion of the drift waves and their growth is determined by the species of particle with the highest curvature drift rate, which are the beam or `primary' particles.

\subsection{Curvature drift instability}

Particles moving along the curved magnetic field undergo drift orthogonal to the plane of field curvature. It is convenient to define a local Cartesian coordinate system, with the $z$ axis along the direction of the magnetic field and the curvature drift velocity along the $x$-axis. The drift speed is given by
\begin{equation}
u_b=\frac{c\gamma v_z}{\omega_{_B}R_{c}},
\label{ub}
\end{equation}
where $\omega_{B}=eB/mc$, $R_{c}$ is the radius of curvature of the magnetic field line, $\gamma$ is the Lorentz factor of the particle, and $v_z\approx c$ is the particle velocity along the magnetic field line. 

Curvature drift can cause low-frequency, nearly transverse drift waves to grow \citep{KMM91a, KMM96}. These waves propagate across the magnetic field, so that the angle $\theta$ between ${\bmath k}$ and ${\bmath B}$ is close to $\pi /2$, corresponding to $k_{\perp}/k_z\gg 1$, with
$k_{\perp} = (k_{x}^{2}+k_{y}^{2})^{1/2}$. Assuming $\gamma (\omega /\omega_{_B})\ll 1,\ \left( u_b/c\right) ^{2}\ll1,\ k_z/k_{x}\ll1\ $ and $k_{y}\rightarrow 0$. The dispersion equation for the drift wave may be approximated by  \citep{KMM91b, KMM96}
\begin{equation}
\omega =k_{x}u_{b}+k_zv_z+i\Gamma,
\label{dr1}
\end{equation}
with $u_{b}$ given by (\ref{ub}) and where $\Gamma$ corresponds to the growth rate. The growth rate is largest for
\begin{equation}
k_{\perp }^{2}\ll \frac{\omega _{p}^{2}}{\gamma _{p}^{3}c^{2}},
\label{10}
\end{equation}
where $\omega _{p}$ is the plasma frequency. The maximum growth rate is given approximately by
\begin{equation}
\Gamma \simeq \left( \frac{n_{b}}{n_{p}}\right)^{1/2}
\left(\frac{\gamma_{p}^3}{\gamma_b}\right)^{1/2} k_{x}u_{b}.
\end{equation}

The growing drift waves draw their energy from the motion of the beam particles along the magnetic field lines. The resonance condition that allows their growth requires $k_{x}u_{b}\neq 0$. That is, the drift motion of the beam particles is essential for the resonance condition to be satisfied, but the drift motion does not provide the free energy that drives the instability. Assuming $1/k_x$ is the order magnitude of the transverse
dimension of the magnetosphere (cf. eq 7), from (1) one has $k_xu_b
\sim  c^2\gamma r/\omega_{B*}R^3_*\approx0.053(r/10^8\,{\rm cm})\,{\rm s}^{-1}$,
where $\gamma\sim10^7$, $R_*=10^6\,\rm cm$ is the star's radius, $r$ is the emission 
radius, and $\omega_{B*}$ is the cyclotron frequency with the magnetic field 
$B_*$ taken on the polar cap. The growth rate of the drift wave is rather small,
i.e. generally $\Gamma<\Omega$,  but even a small growth rate can be effective because the waves propagate nearly perpendicular to the magnetic field and can stay in resonance for a long time $\gg 2\pi/\Omega$. Each particle gives up only a small part of its energy to the waves as it propagates through an interaction region where the waves are growing.
New particles continually pass through this interaction region where they spend a relatively short time so that the back reaction of the waves upon the particles is unimportant. Thus the energy accumulates in the waves without quasi-linear saturation. The wave amplitude grows until nonlinear processes
redistribute the wave energy.

\subsection{Nonlinear evolution of drift waves}

Wave-wave interactions include three-wave and four-wave couplings and induced scattering. A three-wave interaction involves the quadratically nonlinear current which is an odd function of the sign of the charge, $\propto e^3$, so that electrons and positrons contribute with opposite sign, making this effect relatively weak in a pulsar plasma \citep{LM94}. Four-wave interaction, which involve the cubic nonlinearity, $\propto e^4$, are intrinsically small, and tend to be important only when three-wave processes are weak or forbidden. Induced scattering may be regarded as a nonlinear absorption process, absorbing the beam between the scattered and unscattered waves. This leads to a transfer of waves from higher to lower frequencies, with the rate proportional to the product of the wave energies in the scattered and unscattered waves. Induced scattering should be the dominant nonlinear process in the present case.

Induced scattering occurs when the resonant condition
\begin{equation}
\omega-\omega^\prime-(k_z-k_z^\prime)v_z=0,
\label{resonant}
\end{equation}
is satisfied. Note that induced scattering is dominated by the pairs, and not by the
primary particles, so that we may estimate $|v_z|\approx1-1/2\gamma_p^2$.
Substituting (\ref{dr1}) into (\ref{resonant}) one obtains the ratio
\begin{equation}
\frac{k_x-k_x^\prime}{k_z-k_z^\prime}=- \frac{c}{u_b}
\frac{1}{2\gamma_p^2}.
\label{13}
\end{equation}
We estimate this ratio be be in the range $10$--$100$. The positive sign of $k_z$ corresponds to waves propagating from the pulsar to the light cylinder.

This induced scattering transfers waves from higher to slightly lower frequencies, in a way analogous to the the transfer of Langmuir waves in an isotropic thermal plasma \citep{TS,GS,G83,M86}. The drift waves generated by the instability have $k_z/k_x \ll 1$, and the nonlinear transfer is from larger to smaller $|k_x|$. From (\ref{13}) it follows that during each scattering process the changes in the wave numbers satisfy $|\Delta k_x|\gg|\Delta k_z|$. This implies that the transfer of energy through the scales should be determined by $k_x$ alone and that the process is closely analogous to the induced scattering of Langmuir waves in an isotropic thermal plasma. As the transfer to smaller $k_x$ the waves move into a range of $k_x$ where there is no effective wave dissipation, due to the resonance condition not being satisfied for any particles. This energy transfer continues until the length scale approaches the transverse dimension, $d$, of the magnetosphere corresponding to a minimum $k_x^{\rm min}\sim1/d$. We identify $d$ are the radius of a circle about the magnetic pole, and express it in terms of the emission radius, $r$, which we write as a fraction, $\xi\ll1$, of the light cylinder radius, $R_{\rm LC}$. This gives
\begin{equation}
d  \approx\xi^{3/2}R_{\rm LC}=(1.6\times10^8\,{\rm cm})
 \left({\xi\over0.1}\right)^{3/2}\left({P\over{1\rm\,s}}\right).
 \label{d}
\end{equation} 
For such small $k_x$, the treatment in terms of plane waves becomes inappropriate.

When the wavelength of the drift waves in the azimuthal direction is of order $d$, the waves need to be treated in terms of azimuthal harmonics, $\propto\exp(im\phi)$, rather than plane waves. Qualitatively, the important point is that relevant waves have an integral number, $m$,  of nodes and anti-nodes around the circumference of the circle. The wave energy accumulates in this wavelength region provided that the rate of nonlinear pumping into the region with $k_x \sim k_x^{\rm min}$ is
sufficiently high. Specifically, the nonlinear transfer must occur at a faster rate than the drift waves propagate out of the region, allowing the `reservoir' of energy (in wave number space) at
$k_x \approx 10^{-8} \,{\rm cm^{-1}}$ to be maintained.  The propagation time for waves with $k_z/k_x
\approx 10^{-2}$ is $\tau\sim (\xi/\Omega)(k_x/k_z)\approx 2(\xi/0.1)(P/1\,{\rm s})\,\rm s$. Consequently, the reservoir is exhausted in time $\tau$. We need to compare this with the characteristic time for the nonlinear transfer to $k_x \sim k_x^{\rm min}$.

The nonlinear transfer rate due to induced scattering may be estimated as
\begin{equation}
\Gamma_{NL} \approx \eta
\frac{\omega_p^2}{\omega \gamma_p^2},
\label{GammaNL}
\end{equation}
where $\eta=U/n_pm_ec^2\gamma_p$ is the ratio of the energy density of the waves, $U$, to the energy density in the beam particles. It follows from (\ref{GammaNL}) that setting $\Gamma_{NL}$ equal to the linear growth rate implies only a modest energy density in the waves, $\eta\ll 1$. It then follows that the transfer time, $1/\Gamma_{NL}$, for the drift waves to small $k_x$ is of the order of the time $\tau$. We assume that a balance is established between input of wave energy due to the instability at relatively large $k_x$ and nonlinear transfer of energy to smaller $k_x$. 

We conclude that it is plausible that drift waves do accumulate in the long-wavelength region. The nonlinear transfer should favour one specific eigenmode becoming dominant, corresponding to a specific azimuthal wave pattern, that is, a pattern with a specific integer number, $m$, of wavelengths  around a circle centered on the magnetic pole.

\subsection{Frequency of the drift eigenmodes}

A remarkable feature of this model for the generation of low-frequency drift waves is that the resulting frequency of the waves is of the same order of magnitude as the rotation frequency of the star. In particular, the ratio of the wave frequency to the rotation frequency depends only on the mode number, $m$, and the ratio, $\xi=r/R_{\rm LC}$, of the emission radius to the light cylinder radius. This occurs because the Lorentz factor of the beam particles is limited by the potential available, which is proportional to $B_*\Omega^2$, so that the curvature drift speed (\ref{ub}) is independent of the strength of the magnetic field, $B_*$. Let $\gamma_{\rm max}$ be the maximum Lorentz factor, corresponding to the maximum potential energy available (between the magnetic axis and the last closed field lines) in a vacuum model. One has $\gamma_{\rm max}\approx\Omega^2\omega_{B*}R_*^3/c^3$, and with $R_c\approx\xi^{1/2}R_{\rm LC}$ this implies
\begin{equation}
{u_b\over c}\approx{\gamma\over\gamma_{\rm max}}\,\xi^{5/2},
\label{ubc}
\end{equation}
Then for $k_x^{\rm min}=m/d$ with $d$ given by (\ref{d}), the ratio of the frequencies becomes
\begin{equation}
{\omega_m\over\Omega}={k_x^{\rm min}u_b\over\Omega}
\approx m{\gamma\over\gamma_{\rm max}}\,\xi.
\label{ratio}
\end{equation}
The factor $\gamma/\gamma_{\rm max}$ for the primary particles depends on the details of the model. Nevertheless, it is evident that for plausible values of $m\sim20$ and $\xi\sim0.1$ the two frequencies are of the same order of magnitude. This is an important feature of our model in that it provides a natural explanation of the period, $P_3$, associated with drifting subpulses, as discussed in the next section.

The estimate (\ref{ratio}) is derived assuming that the term $k_zv_z\approx k_z c$ is no larger than $k_x^{\rm min}u_b$. A (relativistic) particle is accelerated along the magnetic field due to the electric field of the drift wave, $\delta E_\parallel$, and $k_z c<\omega_m$ implies that this acceleration is temporally limited rather than spatially limited. 

\subsection{Amplitude of the drift waves}

The effect of the drift waves on the particle distribution that is responsible for the radio emission may be estimated by considering the amplitude, $\delta\gamma$, of the oscillating part of the Lorentz factor of an electron or a positron due to the forced motion in the wave. Let $\delta B$ be the magnetic amplitude of the wave. Then the relation between $\delta\gamma$ and $\delta B$ may be determined as follows.

The oscillations of a particle in a wave may be either approximately spatial or approximately temporal, depending on whether the phase velocity of the wave is much less or much greater, respectively, than the velocity of the particle. In the present case, the relevant motion is along the magnetic field lines and the relevant phase velocity is $\omega_m/k_z$. The dispersion relation $\omega_m=k_xu_b$ and the assumption $k_z c\ll\omega$ then imply
\begin{equation}
\delta E_\parallel=u_b\,\delta B,
\qquad
\delta\gamma\approx {e\delta E_\parallel\over\omega_mm_ec},
\label{deltaB1}
\end{equation}
where $\omega_p$ is the plasma frequency for the pairs. The amplitude $\delta B$ may be expressed in terms of the ratio $\eta$, introduced in (\ref{GammaNL}). One has
\begin{equation}
\eta={\delta B^2\over\mu_0n_pm_ec^2\gamma_p}
={\delta B^2\over B^2}{\omega_B^2\over\omega_p^2}{1\over\gamma_p},
\label{deltaB2}
\end{equation}
where $\omega_p$ is the plasma frequency for the pairs. The amplitude of the oscillations in the Lorentz factor may then be written in the form
\begin{equation}
{\delta\gamma\over\gamma_p}=\left({\eta\over\gamma_p}\right)^{1/2} {\omega_p\over\omega_m}
{u_b\over c},
\label{deltaB3}
\end{equation} 
with $u_b/c$ given by (\ref{ubc}). The assumption that the waves have small amplitudes requires not only $\eta\ll1$ but also $\delta\gamma/\gamma_p<1$. The drift waves modulate the radio emission only if $\delta\gamma/\gamma_p$ is large enough to modify the emission process, as discussed below. 

\section{Drifting subpulse phenomenon}

A model for subpulse drift based on long-wavelength drift waves has a number of attractive features. As already noted, the waves should accumulate to form a specific azimuthal pattern,  qualitatively consistent with the carousel model. Moreover, the fact that the frequency of the waves is comparable with the rotation frequency of the star implies that this pattern drifts in azimuth, again consistent with the observed drifting patterns, and providing a natural explanation for $P_3$. Another favourable feature of the drift-wave model  is that the electric and magnetic fields of the waves provide a natural mechanism for modulating the radio emission. 

\subsection{Modulation of the emission process}

Any model for pulsar radio emission must involve some form of coherent emission, and most emission mechanisms \citep{M95}  involve a maser-type or reactive instability.  The most widely favoured emission processes involve a beam instability generating waves that indirectly produce the escaping radiation. There are also several maser emission processes: maser curvature emission \citep{LM92,LM95}, linear acceleration emission \citep{M78,R95} and free-electron maser emission \citep{FK04}. A common feature of all such instability mechanisms is that they are driven by free energy associated with a peak in the momentum distribution. Specifically, if $f(p)$ is the one-dimensional momentum distribution function of the particles, these instabilities are driven by particles with $df(p)/dp>0$. Such a rising portion of the distribution function is confined to the lowest energies, below the maximum in $f(p)$. Typically the growth rate for these instabilities decreases with increasing Lorentz factor at the peak in the distribution function. Recent estimates suggest a relatively modest value for the peak Lorentz factor, $\gamma_p\sim10$--100 \citep{ZH00,HA01,AE02}. A modulation $\delta\gamma/\gamma_p\sim0.1$--1 would strongly modify the growth of the waves in any such instability.

The magnetic field in the drift waves also affects the pattern of any radio emission. The magnetic field in the waves is nearly perpendicular to the pulsar magnetic field ${\bf B}$, and leads to as a periodic perturbation of it. This perturbation affects the angle of emission, which is centred on the direction of the magnetic field, causing the emission angle relative to the star to oscillate in the polar direction away from the magnetic axis. Such oscillations imply that the emission pattern from dipolar field lines originating from a fixed magnetic latitude on the star form a pattern with $m$ oscillations in the polar direction about a circle on the sky around the magnetic pole. The angle through which the direction of emission is deflected is $\sim\delta B/B$, cf.\ (\ref{deltaB2}). 

In principle, the oscillations in the magnetic field, $\delta B$, might also act as a wiggler field in a free-electron maser model \citep{FK04}. However, the frequency of the emission would be $\sim k_xc\gamma_p^2$, which is too low to be of observational interest from $k_x\sim k_x^{\rm min}$ and $\gamma_p\sim10$--100 for the bulk of the pairs.

\subsection{Interpretation of $P_3$}

As already noted, the frequency, $\omega_{m}$, of the drift waves is of order the rotation frequency, $\Omega$, and this provides an interpretation of the period $P_3$ in terms of a beat between these frequencies. Suppose that $\omega_m$ is close to an integer, $n$, time the rotation frequency, 
\begin{equation}
\omega_{m}=n\Omega \pm \Delta.
\label{Delta}
\end{equation}
Then the drifting pattern repeats with a period $P_3=2\pi/\Delta$ \citep{CR04}.  For $ \Delta>0$ the subpulse drift is in the direction of the pulsar rotation and for $\Delta<0$, the wave phase falls behind the
pulsar rotation and subpulse drift is in the opposite direction.  After the time $P_3=2\pi/|\Delta|$ the phase returns to the original position and the pattern repeats.

The ratio of the numbers $n$ and $m$ may be estimated as follows. The wave number is $k_x^{\rm min}=m/d$, and the frequency is $\omega_m=mu_b/d$. Assuming a dipolar field, this give
\begin{equation}
\frac{n}{m}=\frac{\gamma}{\gamma_{\rm max}}\,\xi,
\label{nm}
\end{equation}
where (\ref{ubc}) and (\ref{Delta}) are used. The ratio (\ref{nm}) for emission at a height corresponding to $\xi\sim0.1$ and for  $\gamma/\gamma_{\rm max}\sim1$ is consistent with $n$ of a few and $m\sim20$. Such numbers can plausibly explain the main features of subpulse drifting.

\subsection{Further implications of the model}

Our model differs from all other models for subpulse drifting in that it relies purely on magnetospheric oscillations and not on effects related to the surface of the star. In our discussion we have assumed that the emission height is a fraction $\xi\sim0.1$ of the distance to the light cylinder. Although the results are not particularly sensitive to this choice of $\xi$, it is clear that the model cannot explain radio emission originating from close to the surface of the star, e.g.~\cite{KG98}. The purely magnetospheric character of the modulation of the radio emission is an attractive feature of the model, but it does imply that the source of the radio emission is relatively high in the magnetosphere. In particular, it is not consistent with some recent models for radio emission from near the surface of the star, cf.\ \cite{GLM04}. The model also provides scope for interpreting a variety of phenomena. We comment briefly on some possibilities.

Pulsars PSR 2303+30 and PSR 1944+17 exhibit a complicated picture of drifting subpulses that is usually explained by the existence of two or more drift systems \citep{MT}. An interpretation within the framework of the present model is that two different azimuthal harmonics (two different $m$ values) for the drift waves are excited with approximately the same amplitude. The phenomenon of nulling is associated with drifting subpulses, and one possible interpretation is that emission is possible only in the presence of the drift waves (occurring when the oscillations in $\gamma_p$ have their minimum value), and that for some reason the pattern of drift waves disappears. The transient features observed as the drifting subpulses are re-established following a null \citep{vLetal03} may be interpreted in terms of the re-establishment of the drift-wave pattern. Although these suggestions are only speculative, the important qualitative point is that the drift-wave model provides a purely magnetospheric framework for interpreting phenomena associated with drifting subpulses, without any reference to surface phenomena on the star.

\section{Conclusions}

In this paper we attribute the subpulse drifting phenomenon to modulation of the radio wave generation region by drift waves. Relatively small scale drift waves are generated by a beam instability due to the curvature drift of `primary' particles. Nonlinear processes, specifically induced scattering of the drift waves by plasma particles, transfers the drift wave energy to longer wavelengths. At sufficiently long wavelengths the geometry of the magnetosphere becomes important, and the eigenmodes correspond to specific azimuthal numbers $m$. We argue that the nonlinear wave processes should favour one eigenmode becoming dominant, so that the wave energy accumulate in waves with this $m$. The waves then form a pattern with an integral number, $m$, of nodes around a circle centre on the magnetic pole. We suggest that the $m$ nodes or anti-nodes around a circle correspond to the number of subsources in the carousel model. This pattern varies in time at the wave frequency, and the mismatch, $\Delta=\omega-n\Omega$, between the wave frequency and the nearest harmonic of the rotation frequency determines the period, $P_3=2\pi/\Delta$, at which the pattern repeats. For the model to account for the main features of the observed subpulse drift requires that $n=\omega/\Omega$ be of order unity, and $m$ be of order 20. For plausible parameters we find $m/n$ of the required order, cf.\ (\ref{nm}).

These long-wavelength drift waves can affect the radio wave generation process by modulating the low-energy portion of the particle distribution function due to acceleration along the magnetic field lines by the electric field of the drift waves. Most favoured emission processes involve an instability driven by the rising portion, $df(p)/dp>0$, of the distribution function, and the emission processes are sensitive to modifications in this low-energy portion of the distribution function. The magnetic field in the drift waves also modifies the local direction of the magnetic field lines, leading to a change in the angle of emission relative to the star. 

Qualitatively, perhaps the most important feature of the drift wave model is that it is purely magnetospheric, and is not related to processes on the surface of the star. We envisage emission from a height that is a fraction, $\xi$, of the light-cylinder radius, with $\xi\sim0.1$ in our numerical estimates, and the drift-wave pattern is assumed to be set up at this height. Our favored suggestion for modulation of the emission process, to produce drifting subpulses, is that the parallel electric field in the drift waves modifies the distribution of particles, and allows maser emission to be effective only in $m$ localized regions, implying a carousel-like model. Another possibility is that the effect of the drift waves is to modulate the angle of emission so that it oscillates in and out of sight as the carousel rotates. 

\section*{Acknowledgments}
We thank Simon Johnston for helpful comments on the manuscript.

\end{document}